\documentclass[aps,prl]{revtex4-2}

\usepackage{eurosym}
\usepackage{amsmath,amssymb}
\usepackage{graphicx}
\usepackage{dcolumn}
\usepackage{bm}
\usepackage{color}
\usepackage{multirow}
\usepackage{float}
\usepackage{array}
\usepackage{dcolumn}
\usepackage{booktabs}
\usepackage[detect-all]{siunitx}
\usepackage{makecell}
\usepackage{xcolor}
\usepackage{yfonts}
\usepackage[normalem]{ulem}
\newcolumntype{P}[1]{>{\centering\arraybackslash}p{#1}}
\begin{document}

\title{Possible Existence of  $^3_\phi$H, $^4_\phi$H, $^4_\phi$He, and $^5_\phi$He Nuclei
 }
\author{R. Lazauskas$^{1}$, R. Ya. Kezerashvili$^{2,3,4}$, I. Filikhin$^{5}$}
\affiliation{$^{1}$IPHC, CNRS/IN2P3, Universit´e de Strasbourg, 67037 Strasbourg, France\\
$^{2}$New York City College of Technology, The City University of New York,
Brooklyn, NY, USA\\
$^{3}$The Graduate School and University Center, The City University of New
York, New York, NY, USA\\
$^{4}$Long Island University, Brooklyn, NY, USA\\
$^{5}$North Carolina Central University, Durham, NC, USA \\
}




\begin{abstract}
\noindent 
Motivated by recent HAL QCD simulations of the $\phi N$ interaction in the $^4S_{3/2}$ channel and its modification in the $^2S_{1/2}$ channel, we develop a first-principles few-body framework that embeds these potentials into configuration-space Faddeev--Yakubovsky equations. We predict bound $^4_\phi\mathrm{H}$, $^4_\phi\mathrm{He}$, and $^5_\phi\mathrm{He}$ nuclei by performing calculations for $\phi$-mesic $\phi NNN$ and $\phi NNNN$ systems. Both spin-dependent and spin-independent $\phi N$ interactions are considered, leading to deeply and moderately bound states, respectively. The deeply bound states originate from the strong attraction in the $^2S_{1/2}$ $\phi N$ channel. Coulomb shifts of the binding energies are evaluated.
Our findings provide the binding mechanism and demonstrate the importance of short-range $\phi N$ attraction.
\end{abstract}

\maketitle
\date{\today }

\textbf{Introduction.}
Recent HAL QCD extractions of the $\phi N$ interaction for the first time, enabled a rigorous configuration-space Faddeev–Yakubovsky treatment  of $\phi$-mesic few-body systems, grounded in lattice-QCD-constrained dynamics. This advancement builds the opportunity in the details investigation of the $\phi$-mesic $\phi NNN$ and $\phi NNNN$ systems.

Motivated by the strong antikaon–nucleon attraction, the meson–baryon–baryon dibaryon $\bar{K}NN$ was predicted in 2002~\cite{Kpp}, triggering extensive theoretical and experimental studies of $\bar{K}$-nuclear few-body systems (see reviews~\cite{Hyodo2012,Gal2016,Kez10,Kez20}). These investigations employed a broad range of few-body techniques, including AGS equations~\cite{Alt1967,Marri2016a,Marri2016b,Marri2019,Shevchenko2022},  Faddeev equations in momentum~\cite{Shevchenko2007a,Shevchenko2007b,Ikeda2007,Ikeda2009,Ikeda2010,Revai2014} and configuration~\cite{Kez2017,Maeda2013} spaces, fixed-center approximations~\cite{Bayar2011,Bayar2012,Bayar2013}. In contrast, few-body systems containing a $\phi$ meson and multiple nucleons remain largely unexplored, despite growing interest in $\phi$-mesic nuclei and the long-standing question of a possible $\phi N$ bound state. Recent progress has been driven by the ALICE measurement of the $p$–$\phi$ correlation function~\cite{ALICE2021} and the derivation of the $\phi N$ potential from near-physical $(2+1)$-flavor lattice QCD by the HAL QCD collaboration~\cite{Lyu22}. A subsequent reanalysis~\cite{Chizzali2024} disentangled the spin-$(\tfrac{1}{2},\tfrac{3}{2})$ components, indicating a possible bound state in the $\phi N(^2S_{1/2})$ channel, with absorptive effects arising from $S$-wave coupling to $\Lambda K$ and $\Sigma K$ channels~\cite{Lyu22,Chizzali2024}.

Three-body calculations of the $\phi NN$ system have been carried out by solving differential Faddeev equations~\cite{BSS,FKVPRD2024}, variational~\cite{Bel2008}, hyperspherical expansions~\cite{EA24}, and coupled-channel complex-scaling approaches~\cite{Wen2025}, employing different $\phi N$ interactions~\cite{G2001,Lyu22,Chizzali2024}. The resulting binding energies vary widely, reflecting strong model dependence. However, $\phi$-mesic systems with four and five particles, $\phi NNN$ and $\phi NNNN$, have not been studied to date. 

In this Letter, we report the first systematic few-body study of $\phi$-mesic systems up to five particles. Our approach integrates HAL QCD–derived $\phi N$ interactions~\cite{Lyu22,Chizzali2024} combined with the MT I–III $NN$ potential~\cite{Malfliet1969,MTcorr}. Faddeev–Yakubovsky equations~\cite{Yakubovsky1967} for two distinct particle species were formulated in configuration-space and solved numerically, extending earlier four- and five-body frameworks~\cite{Nogga_Phd,nogga2002hypernuclei,Lazauskas2018a,Lazauskas2018b}. This approach enables the first systematic and unified analysis of $\phi NNN$ and $\phi NNNN$ bound states, constrained by lattice QCD dynamics.

\textbf{Interaction Potentials.} We study $\phi$-mesic systems $\phi NN$, $\phi NNN$, and $\phi NNNN$ within a nonrelativistic potential framework using $\phi N$ and $NN$ interactions. The $\phi N$ interaction is defined in $^{4}S_{3/2}$ and $^{2}S_{1/2}$ channels. The $\phi N(^{4}S_{3/2})$  interaction is obtained using recent experimental measurements of the $p$–$\phi$ correlation function by ALICE~\cite{ALICE2021} and first-principles $(2+1)$-flavor lattice QCD calculations employing the HAL QCD method~\cite{Lyu22}. 
The $\phi N(^{2}S_{1/2})$ potential
was constrained by a spin-resolved reanalysis~\cite{Chizzali2024} used recent experimental measurements of the $p$–$\phi$ correlation function by ALICE~\cite{ALICE2021}. These developments provide, for the first time, lattice-QCD-constrained inputs suitable for rigorous few-body calculations.

Following~\cite{Lyu22,Chizzali2024}, the $S$-wave $\phi N$ potentials for $^2S_{1/2}$ and $^4S_{3/2}$ channels can be written as a sum of short-range Gaussians and a long-range two-pion–exchange tail,
\begin{equation}
V_{\phi N}(r)=\beta\left(a_1 e^{-r^2/b_1^2}+a_2 e^{-r^2/b_2^2}\right)
+a_3 m_\pi^4 F(r,b_3)\left(\frac{e^{-m_\pi r}}{r}\right)^2 ,
\label{HALQCD}
\end{equation}
with an Argonne-type formfactor $F(r,b_3)=(1-e^{-r^2/b_3^2})^2$~\cite{Wiringa95}.
Using central values from~\cite{Lyu22}, the parameters are
$a_1=-371$ MeV, $a_2=-119$ MeV, $b_1=0.13$ fm, $b_2=0.30$ fm, $b_3=0.63$ fm, and $a_3 m_\pi^4=-97$ MeV·fm$^2$ and QCD masses $m_\pi=146.4$ MeV, $m_N=954$ MeV, and $m_\phi=1048$ MeV.

For $\beta=1$, one obtains the HALL QCD potential \cite{Lyu22} that describes the $\phi N({}^4S_{3/2})$ channel. This potential is attractive at all distances but does not support bound states in either $\phi N$ or $\phi NN$ systems~\cite{Lyu22,EA24,FKVPRD2024}. Constraining this channel with lattice-QCD scattering parameters~\cite{Lyu22} and fitting ALICE data, Ref.~\cite{Chizzali2024} inferred a strongly enhanced $\phi N({}^2S_{1/2})$ interaction with $\beta =$6.9$_{-0.5}^{+0.9}$(stat.)$_{-0.1}^{+0.2}$(syst.), generating a $\phi N$ bound state. The long-range tail is identical in both spin channels and dominated by scalar–isoscalar two-pion exchange~\cite{Kreinm4}. 

For the nucleon–nucleon interaction, we employ the MT I–III potential~\cite{Malfliet1969,MTcorr},
$V_{NN}^s(r)=v_r\frac{e^{-\mu_r r}}{r}+v_a^s\frac{e^{-\mu_a r}}{r},$
with $\mu_r=3.11$ fm$^{-1}$, $\mu_a=1.55$ fm$^{-1}$, and $v_r=1438.72$ MeV·fm.
Standard strengths are $v_a^{s=0}=-513.968$ and $v_a^{s=1}=-626.885$ MeV·fm, which allow for reproducing fairly well the deuteron binding energy and $np$ scattering lengths using physical masses. For calculations employing QCD-motivated nucleon masses we readjust to 
$v_a^{s=0}=-510.493$ MeV·fm and $v_a^{s=1}=-623.269$ MeV·fm, yielding $B_d=2.2246$ MeV and $a_{np}=23.737$ fm.


\textbf{Theoretical frameworks.} The Faddeev–Yakubovsky equations (FYE)~\cite{Yakubovsky1967} provide a rigorous and systematic framework for nonrelativistic few-body calculations. Their decomposition of the wave function into all cluster and subcluster partitions ensures an unambiguous identification of genuine bound states, while their structure is particularly efficient when the interaction is restricted to a limited number of partial waves. This makes the FYE especially well suited for $\phi$-mesic systems, where the $\phi N$ interaction exhibits strong spin dependence and contributions beyond the $S$ wave are presently unconstrained and can be neglected.

Derivation of FYE starts by defining Faddeev components (FC)
\begin{equation}
\psi_{ij}= (E-\hat{H}_{0}) ^{-1}\hat{V}_{ij}  \Psi,
\end{equation} 
where $\hat{H}_0$ is a free Hamiltonian of the system and $\hat{V}_{ij}$ is interaction between particles $i$ and $j$. The systems wave-function can be expressed in terms of FC as:
$\Psi=\sum_{i<j}\psi_{ij}$.
 The number of FC $\psi_{ij}$ equals the number of binary interactions. A former equations constitutes a set of three Faddeev equations for a three-body problem.
\begin{figure}[h!]
\begin{center}
\includegraphics[width=10.pc]{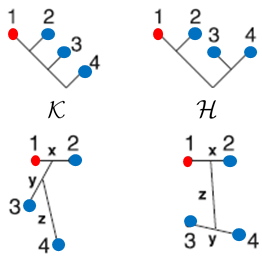}
($a$)
\includegraphics[width=25.pc]{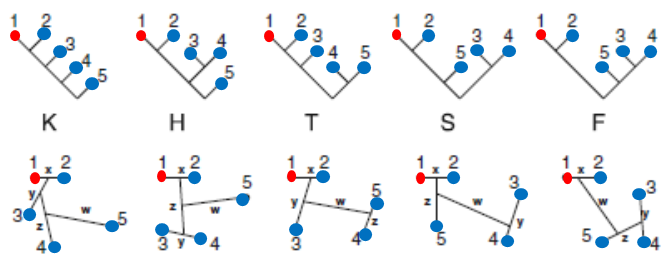}
($b$)
\end{center}
\caption{ 
The upper ($a$) and ($b$) panels illustrate the distinct topological decompositions of the fully interacting $\phi NNN$ and $\phi NNNN$ clusters, respectively, into their substructures.
The lower panel presents the Jacobi coordinate systems associated with each of these cluster partitions. Low panels: ($a$) K-like and H-like components, as $\Phi^l_{12,3}$ and $\Phi^l_{12,34}$, respectively. Permuting the particle indices yields 12 $\mathcal{K}$-type components and 6 $\mathcal{H}$-like components; ($b$) The five independent 5-body FYE components are denoted as K, H, T, S, F. The permutation of particle indexes gives 60 K-type amplitudes and 30 for each H, T, S, and F-type amplitudes \cite{Lazauskas2018a,Lazauskas2018b}. Ultimately, this yields a system of 180 Faddeev-Yakubovsky equations.} 
\label{Yakub4}
\end{figure}
For systems with more than three particles, the three-body Faddeev components can be further decomposed into four-body components using the Yakubovsky procedure~\cite{Yakubovsky1967}. The four-body Faddeev-Yakubovsky components (FYC) are defined as
\begin{equation}
\mathcal{K}_{ij,k}= (E-\hat{H}_{0}-\hat{V}_{ij} )^{-1}(\psi_{ik}+\psi_{jk}), \qquad 
\mathcal{H}_{ij}^{kl}= (E-\hat{H}_{0}-\hat{V}_{ij} )^{-1}\psi_{kl}\label{eq_4bH},
\end{equation}
with the reconstruction relation
\begin{equation}
\psi_{ij}=\mathcal{K}_{ij,k}+\mathcal{K}_{ij,l}+\mathcal{H}_{ij}^{kl}
\end{equation}
and depicted in panel (a) of Fig.~\ref{Yakub4}. For a four-body system ${\mathcal{K}}$-type components contain the asymptotes of the 3+1 cluster channels, whereas ${\mathcal{H}}$-type components contain the 2+2 asymptotes.
By permuting particle indices
$(ijkl)$, one obtains 12 distinct ${\mathcal{K}}$-type components and 6 ${\mathcal{H}}$-type components, coupled 
through a set of 18 differential equations~(\ref{eq_4bH}). This set of 18 equations/components constitute a system of 4-body FYE.  
For the five-body problem, the four-body components can be further decomposed iteratively, as shown in panel (b) of Fig.~\ref{Yakub4}. Five distinct types of five-body components can be defined~\cite{Sasakawa1977,Lazauskas2018a}
\begin{eqnarray}
\textsl{\textbf{K}}_{ij,k}^{l}&=&(E-\hat{H}_{0}-\hat{V}_{ij}) ^{-1}\hat{V}_{ij}\left(\textsl{\textbf{K}}_{ik,j}^{l}+\textsl{\textbf{K}}_{jk,i}^{l}+\mathcal{K}_{ik}^{l}+\mathcal{K}_{jk}^{l}+\mathcal{H} _{ik}^{jl}+\mathcal{H}
_{jk}^{il}\right),   \\
\textsf{\textbf{H}}_{ij}^{kl}&=&(E-\hat{H}_{0}-\hat{V}_{ij}) ^{-1}\hat{V}_{ij}\left(\textsf{\textbf{H}}_{kl}^{ij}+\mathcal{K}
_{kl,j}+\mathcal{K} _{kl,i}\right),  \\
\textsf{\textbf{T}}_{ij,k}^{lm}&=&(E-\hat{H}_{0}-\hat{V}_{ij}) ^{-1}\hat{V}_{ij}\left(\textsf{\textbf{T}}_{ik,j}^{lm}+\textsf{\textbf{T}}_{jk,i}^{lm}+\mathcal{H}_{ik}^{lm}+\mathcal{H} _{jk}^{lm}\right),  \\
\textsf{\textbf{S}}_{ij}^{lm}&=&(E-\hat{H}_{0}-\hat{V}_{ij}) ^{-1}\hat{V}_{ij}\left(\textsf{\textbf{F}}_{lm}^{ij}+\mathcal{H}_{lm}^{jk}+\mathcal{H}_{lm}^{ik}\right),  \\
\textsf{\textbf{F}}_{ij}^{lm}&=&(E-\hat{H}_{0}-\hat{V}_{ij}) ^{-1}\hat{V}_{ij}\left(\textsf{\textbf{S}}_{lm}^{ij}+\mathcal{K}_{lm,k}\right)
\end{eqnarray}%
The five-body components are related to the four-body ones through
\begin{equation}
\mathcal{K}_{ij,k}=\textsl{\textbf{K}}_{ij,k}^{l}+\textsl{\textbf{K}}_{ij,k}^{m}+\textsl{\textbf{T}}_{ij,k},   \qquad
\mathcal{H} _{ij}^{kl}=\textsl{\textbf{H}}_{ij,kl}+\textsl{\textbf{S}}_{ij,kl}+\textsl{\textbf{F}}_{ij,kl}. %
\end{equation}
In total, the five-body FYE system consists of 180 components/equations: 60 of \textsl{\textbf{K}}-type, and 30 of each $\textsl{\textbf{H-,T-,S-}}$ and $\textsl{\textbf{F}}-$   type, obtained by permuting the five particle indexes $(ijklm)$.

Using the isospin formalism, neutrons and protons can be treated as identical nucleons that differ only in the projection of their isospin quantum number. This particle-permutation symmetry reduces the number of independent FYE components and, consequently, the number of coupled differential equations to solve. In this work, we consider $\phi$-mesic systems: systems of $A$-particles containing one particle that differs from the remaining $A-1$ identical nucleons. 
\begin{table}
    \centering
    \begin{tabular}{ccc}\hline
           $A$   & Number of Eqs. & FC and FYC \\ \hline
       3  &  2& $\psi_{\phi N},\psi_{NN}$\\
       4  &  5 &  $\mathcal{K}_{\phi N,N}^{N},\mathcal{K}_{N N,\phi}^{N},\mathcal{K}_{N N,N}^{\phi},\mathcal{H}_{\phi N}^{NN},\mathcal{H}_{NN}^{\phi N}$\\     
       5  &  16 &   $\textsf{\textbf{K}}_{\phi N,N}^{N},\textsf{\textbf{K}}_{N N,\phi}^{N},\textsf{\textbf{K}}_{N N,N}^{\phi},\textsf{\textbf{K}}_{N N,N}^{N}$        \\      
       & & $\textsf{\textbf{H}}_{\phi N}^{NN},\textsf{\textbf{H}}_{N N}^{\phi N},\textsf{\textbf{H}}_{NN}^{NN},\textsf{\textbf{T}}_{\phi N,N}^{NN},\textsf{\textbf{T}}_{N N,\phi}^{NN},\textsf{\textbf{T}}_{N N,N}^{\phi N}$  \\ 
       &  & $\textsf{\textbf{S}}_{\phi N}^{NN},\textsf{\textbf{S}}_{N N}^{\phi N},\textsf{\textbf{S}}_{NN}^{NN},\textsf{\textbf{F}}_{\phi N}^{NN},\textsf{\textbf{F}}_{N N}^{\phi N},\textsf{\textbf{F}}_{NN}^{NN}$ \\ \hline
    \end{tabular}
    \caption{Number of untrivial Faddeev-Yakubovsky equations and components to consider when solving $A=3$ to $A=5$ body problems for $\phi$-mesic systems.}
    \label{tab:eqs}
\end{table}
Thus, one can reduce the number of independent FYE components to 2, 5, and 16, when solving $A=3$, 4 and 5 particle problems, respectively. The corresponding numbers of independent equations are summarized in Table \ref{tab:eqs}. 
We emphasize that the developed theoretical framework enables rigorous few-body calculations of light hypernuclei containing three or four nucleons bound to baryons such as $\Lambda$, $\Omega$, and $\Xi$. In particular, it allows for a systematic investigation of the $YNNN$ and $YNNNN$ ($Y=\Lambda$, $\Omega$, $\Omega_{ccc}$, and $\Xi$) systems within a fully microscopic few-body approach. 
In the case of $YNNNN$ systems, the applicability of the $Y\alpha$  cluster models can be assessed through direct five-particle calculations.

\textbf{Numerical results and discussion.} To solve the systems of differential equations given above, we employ the numerical methods described in~\cite{K86,La1,La2,La3,KezJPG2024}. The FYC's are expressed in mass-scaled Jacobi coordinates suited to each component’s topology (see Fig.~\ref{Yakub4}).

The spatial, spin, and isospin dependence of the FY component is expressed using the partial-wave expansion. The radial partial amplitudes are expanded in a Lagrange–Laguerre basis using the Lagrange-mesh method~\cite{Baye2015}. The number of basis functions is adjusted according to the partial angular momentum, and reduced systematically for higher partial waves. Since we use S-wave interactions in this work, only amplitudes with $l_x=0$ 
contribute. Other angular momenta were restricted to $max(l_i)<3$, which ensures convergence to four significant digits.

A summary of our results for $A=2$–$5$ $\phi$-mesic systems with physical and QCD-motivated masses is given in Table~\ref{BE345}. Binding is dominated by the strongly attractive $^{2}S_{1/2}$ $\phi N$ interaction, which alone supports a $\phi N$ bound state with several MeV binding energy. In contrast, the $^{4}S_{3/2}$ channel is only weakly attractive and unbound, favoring configurations that maximize the number of $\phi N$ pairs in the $^{2}S_{1/2}$ state. Consequently, $\phi nn$ and $\phi pp$ systems are unbound~\cite{FKVPRD2024,FRV25}, since the Pauli principle forces one nucleon to couple to the $\phi$ in the unfavorable $^{4}S_{3/2}$ channel. For the same reason, binding occurs only for the lowest total isospin states: $T=0$ or $T= 1/2$.

The $\phi np$ system exhibits two bound states with $J^{\pi}=0^-$ and $1^-$, with the $0^-$ ground state favored by simultaneous $^{2}S_{1/2}$ coupling of both nucleons to the $\phi$ meson. The $\phi NNN$ system supports a single bound state with $T=1/2$ and $J^{\pi}=1/2^-$, characterized by a weakly bound third nucleon, $\sim4$ MeV separation energy,  compared to  $\sim18 ~\mathrm{MeV}$ and $\sim34 ~\mathrm{MeV}$ for the first two nucleons. This state remains bound with Coulomb interaction included, implying bound $^4_\phi$H and $^4_\phi$He, whose binding-energy difference, $0.68$ MeV, is comparable to that of $^3$H–$^3$He within the MT I–III model. 
Strong spin dependence of the $\phi N$ interaction leads to significant nucleon pairing effects, producing a deeply bound $T=0$, $J^{\pi}=1^-$ ground state in the $\phi NNNN$ system with a large nucleon separation energy of $\sim23$ MeV. No additional bound states are found for $A=3$–$5$.

The $\phi N$ potential (\ref{HALQCD}) contains a strong short-range attraction and a two-pion–exchange tail at large distances. The strength of the short-range attraction depends on the parameter $\beta$. To gain deeper insight into the dynamics of $\phi$-mesic systems,
we investigated how their properties evolve when the $\phi N$ interaction in the
$^{2}S_{1/2}$ channel is modified.  In Table \ref{BE345} we present results for both $\beta=6.9$ \cite{Chizzali2024} and $\beta=1$. 
For $\beta=1$, the interactions in the  $^{2}S_{1/2}$ and $^{4}S_{3/2}$  $\phi N$ channels are identical, and therefore the binding energies of the $\phi$-mesic systems do not depend on the spin orientation of the $\phi$ meson. In this case, the $T=0$ system, corresponding to $^3_\phi$H nucleus,  has very weakly bound set of three degenerate states with  $J^{\pi}=(0^-,1^-,2^-$), where the $\phi$ is bound to the deuteron by only $34 ~\mathrm{keV}$. Changing $\beta$ value lifts this degeneracy. The binding energy of the $J^{\pi}=2^-$ state is determined solely by the $\phi N$ interaction in $^{4}S_{3/2}$  channel and thus remains independent of  $\beta$. In contrast, the $J^{\pi}=0^-$  is bound purely through the $^{2}S_{1/2}$  interaction. 

Two total-isospin $T=\frac{1}{2}$  bound states exist in the  $\phi NNN$ system, namely $J^{\pi}=\frac{1}{2}^-$ and $J^{\pi}=\frac{3}{2}^-$ ones. These two states are degenerate at $\beta=1$ and correspond to a $\phi$-meson bound inside the trinucleon system: $^3$H or $^3$He. 

The five-body system has only one bound state, $^5_\phi$He, with total isospin $T=0$ and angular momentum $J^{\pi}=1^-$. This state corresponds to $\phi$-meson strongly bound inside the $\alpha$-particle ($^4$He). Nucleon separation energy is largest at $\beta=1$  and, similarly to the lighter systems, decreases once the $\phi N$ subsystem becomes bound, and its binding energy begins to rise.  The  $^4$He nucleus has a resonant excited state -- commonly referred to as the breathing mode of the $\alpha$-particle -- very close to binding \cite{duerinck2025excited}. It is natural to expect a corresponding replica of this state to be bound by adding the $\phi$-meson. This happens for small $\beta$ values, however once $\beta$ is increased this excited state moves below $^4_\phi$H threshold. This tendency is a combination of two effects : on one hand the gap between ground state of $^5_\phi$He system and  $^4_\phi$H threshold gets smaller, on the other hand, the increase in binding energy makes $^5_\phi$He  more compact increasing the contribution of the Coulomb repulsion between the protons in $^5_\phi$He excited state. Absence of stable $A=4$ nuclei other than $^5_\phi$He is the main reason preventing the presence of other  $\phi NNNN$ bound states. 

\begin{table*}[h!]
\caption{\label{BE345}
 Binding energies of light $\phi$-mesic nuclei for $\beta=1$ and $\beta=6.9$. Calculations are performed with physical and QCD-motivated particle masses. UNB denotes an unbound state.  Values in parentheses correspond to results obtained without the Coulomb interaction. All energies are given in MeV and are measured from the corresponding $\phi$-meson separation thresholds.  The core nuclei have the binding  energies $B_d=2.22457$, $B_{^3{\text{H}}}=8.425$, $B_{^3{\text{He}}}=7.763$, $B_{^4{\text{He}}}=29.010$ ($29.805$ without Coulomb) with the QCD-motivated mass of nucleons and  $B_d=2.2304$, $B_{^3{\text{H}}}=8.544$, $B_{^3{\text{He}}}=7.863$, $B_{^4{\text{He}}}=29.536$ ($30.330$  without Coulomb)  with the physical mass of nucleons.}

\centering
\begin{tabular*}{\linewidth}{@{\extracolsep{\fill}} c c c c c c c}
\hline\hline
\rule{0pt}{3ex}
& $^2_\phi$H 
& \multicolumn{2}{c}{$^3_\phi$H} 
& \multicolumn{2}{c}{$^4_\phi$H \ \ \ \ \ \   \ \ \ \ \ \ \ \ \ \ \ \ \ \ $^4_\phi$He} 
& $^5_\phi$He \\[1.0mm]
\cline{2-2}\cline{3-4}\cline{5-6}\cline{7-7}
\rule{0pt}{3ex}
$\beta$
& $J^\pi=\tfrac{1}{2}^-$
& $J^\pi=0^-$
& $J^\pi=1^-$
& $J^\pi=\tfrac{1}{2}^-$
&  $J^\pi=\tfrac{1}{2}^-$
& $J^\pi=1^-$ \\
\noalign{\smallskip}\hline\hline

\multicolumn{7}{c}{QCD-motivated masses} \\
\hline
$1$ & UNB & $0.034+B_d$ & $0.034+B_d$  & $11.64$ & $10.90$ & $40.05\,(40.93)$ \\
$6.9$ & $18.361$ & $51.625$ & $33.132$ & $55.93$ & $55.15$ & $76.77\,(77.57)$ \\\hline

\multicolumn{7}{c}{Physical masses} \\
\hline
$1$ & UNB & $0.157+B_d$ & $0.157+B_d$ & $12.80$& 12.03& $42.57\,(43.43)$ \\ 
$6.9$ & $16.582$ & $50.255$ & $32.134$ & $55.55$ & $54.71$ & $77.91\,(78.72)$ \\ 
\hline

\hline\hline
\end{tabular*}
\end{table*}

\textbf{Concluding remarks.}
We investigated $\phi$-mesic $\phi NN$, $\phi NNN$, and $\phi NNNN$ systems by solving the Faddeev and
Faddeev--Yakubovsky equations using the HAL QCD $\phi N$ interaction.
Although the $\phi N(^{4}S_{3/2})$ interaction is attractive, it does not generate a
$\phi N$ bound state,  whereas the
$\phi N(^{2}S_{1/2})$ interaction supports a $\phi N$ bound state. The spin-dependent interaction enables the
formation of bound  $^3_\phi$H, $^4_\phi$H, $^4_\phi$He, and $^5_\phi$He nuclei. 
Calculations with both physical and QCD-motivated masses consistently confirm
their existence, which can be interpreted as a $\phi$ meson bound inside a
stable nucleus. Thus, we predict the possible existence of bound $\phi$-mesic nuclei $^3_\phi$H, $^4_\phi$H, $^4_\phi$He, and $^5_\phi$He. 

The binding energies are sensitive to the short-range part of the $\phi N$ interaction through the parameter $\beta$. The spin-dependent interaction, which combines the HAL QCD $\phi N({}^{4}S_{3/2})$ potential ($\beta=1$) with a phenomenologically enhanced $\phi N({}^{2}S_{1/2})$ component ($\beta=6.9$), produces deeply bound $\phi$-mesic nuclei due to the strong short-range attraction. In contrast, the spin-independent $\phi N$ interaction with $\beta=1$ yields more moderate binding energies for $^4_\phi\mathrm{H}$, $^4_\phi\mathrm{He}$, and $^5_\phi\mathrm{He}$, while the $\phi NN$ system remains nearly unbound. These results indicate the need for further clarification of the spin dependence of the $\phi N$ interaction. 

Our findings provide new insights into the structure of
$\phi$-mesic nuclei and the role of short-range $\phi N$ interaction. Overall, the present analysis provides a consistent and controlled framework for
describing $\phi$-mesic nuclei within few-body dynamics, 
offers a practical foundation for future studies of heavier
$\phi$-nuclear systems and for exploring the role of hidden strangeness in
nuclear binding. 

\section*{Acknowledgments}
This work is supported by the City University of New York PSC CUNY Research Award No. 68541-00 915 56, and the Department of Energy/National Nuclear Security Administration under Award No. NA0003979 and by French IN2P3 for a theory project “PUMA”. We were granted access to the HPC resources of TGCC/IDRIS under the allocation 2025-AD010506006R3 made by GENCI (Grand Equipement National de Calcul Intensif). 

\bibliography{Bib_Rimas.bib}

\end{document}